# Title: Ultrafast Broadband Strong-Field Tunnelling in Asymmetric Nanogaps for Time-Resolved Nanoscopy

**Authors:** Haoqing Ning[1†], Marios Maimaris[1†], Jiewen Wei[1], Emilie Gérouville[2], Evangelos Moutoulas[2], Zhu Meng[1], Clement Ferchaud[4], Dmitry Maslennikov[1], Navendu Mondal[1], Tong Wang[1], Colin Chow[1], Aleksandar P. Ivanov[3], Joshua B. Edel[3], Saif A. Haque[1], Misha Ivanov[5], Jon P. Marangos[4,] Dimitra G. Georgiadou[2], Artem A. Bakulin[1*]

**Affiliations**:

[1] Department of Chemistry and Centre for Processible Electronics, Imperial College London, UK

[2] Electronics and Computer Science & Optoelectronics Research Centre, University of Southampton, UK

[3] Department of Chemistry, Imperial College London, UK

[4] Department of Physics, Blackett Laboratory, Imperial College London, UK

[5] Max Born Institute, Berlin, Germany

*Email: a.bakulin@imperial.ac.uk
†These authors contributed equally

**Abstract:** Femtosecond-fast and nanometre-size pulses of electrons are emerging as unique probes for ultrafast dynamics at the nanoscale. Presently, such pulses are achievable only in highly sophisticated ultrafast electron microscopes or equally complex setups involving few-cycle-pulsed lasers with stable carrier-envelope phase (CEP) and nanotip probes. Here, we show that the generation of femtosecond pulses of nanoscale tunnelling electrons can be achieved in any ultrafast optical laboratory, using any (deep-UV to mid-IR) femtosecond laser in combination with photosensitive asymmetric nanogap (PAN) diodes fabricated via easy-to-scale adhesion lithography. The dominant mechanism producing tunnelling electrons in PANs is strong-field emission, which is easily achievable without CEP locking or external bias voltage. We employ PANs to demonstrate ultrafast nanoscopy of metal-halide perovskite quantum dots immobilised inside a 10-nm Al/Au nanogap and to characterise laser pulses across the entire optical region (266-6700 nm). Short electron pulses in PANs open the way towards scalable on-chip femtosecond electron measurements and novel design approaches for integrated ultrafast sensing nanodevices.

**Main text:**

**Introduction**

Rapid advances in nanotechnology call for new disruptive approaches to characterise molecular systems and materials with extreme temporal and spatial resolution. Such characterisation approaches can use electrons to probe nanosystems, similarly, to the widely used transmission electron microscopy and scanning tunnelling microscopy (STM). To achieve temporal resolution the electrons should be produced and detected in short bunches, rather than as continuous particle beams.

In the last two decades, the ability of ultrafast lasers to manipulate electrons with high precision in time and space has exhibited remarkable potential for probing molecular dynamics(*1, 2*). Operating within picosecond to attosecond time scales, electron-probe techniques were able to track the processes on the level down to individual atoms and molecules, providing more direct access to previously underexplored issues in material fundamentals and device physics(*3-6*). These issues encompass microscopic ultrafast spin dynamics(*7*), nanoscale current transport(*8*), quantum light source properties(*9*), observation of excitonic states in low-dimensional material(*10, 11*), and the operational mechanisms of individual biological structures(*12*). Efforts in the field of time-resolved electron nanoscopy have primarily focused on developing powerful but bulky and highly sophisticated techniques that integrate commercial electron microscopy instruments with ultrafast lasers(*13, 14*).

Several alternative approaches have been proposed where electrons are generated and/or detected at the nanoscale(*15-17*). For instance, the generation of ultrafast electrical pulses has been demonstrated on metal-insulator-metal structures, such as the tips of STM(*18-20*) and on-chip platforms(*21*). These approaches have been successfully applied to measure nanoscale carrier dynamics in semiconductor materials and molecules with sub-nanometre resolution.

Short nanoscale electron pulses can be generated in the tunnelling nanostructures using the strong (Keldysh parameter $\gamma \leq 1$) alternating electric field of ultrafast optical pulses(*22*). However, when the optical field exhibits temporal symmetry (e.g., for multicycle ultrashort pulses) and irradiate a nanostructure with symmetrical tunnelling barrier, an alternating flow of electrons produces negligible direct current (DC). To address this challenge, various 'photocurrent rectification' strategies have been employed. The most successful was utilising temporally asymmetric optical fields, such as near-IR(*18*) or terahertz(*19*) laser pulses with few optical cycles and controlled carrier-envelope phase (CEP) (*8, 18, 21*). Such methods may provide non-zero net tunnelling currents and even allow modulation of electron pulses' amplitude through CEP control, validating that the timescales of electron-driven processes are comparable with optical cycle. Using CEP stabilized few-cycles sources, sub-femtosecond ultrafast STM with nm spatial resolution has been realised(*23*). An alternative approach to generate pulses of tunnelling current, not requiring sophisticated CEP stabilised apparatus, relied on applying bias voltages to introduce the asymmetry in nanostructures' current-voltage characteristics to achieve the partial rectification of alternating current generated by the optical field(*20, 24*). Unfortunately, this latter approach is prone to low-frequency thermal modulation of dark tunnelling current in the nanostructure(*25-27*) due to plasmonic heating, leading to artefacts and impacting the temporal resolution of the experiments(*28*).

Due to difficulties in working with tunnelling junctions, some researchers applied nanoscale electron emitters like laser-illuminated nanotips(*29-32*) in combination with macroscale electron detection. The conceptual framework surrounding electron emission from nanotips has been extensively developed, with emission probabilities exhibiting similarities to those derived for strong-field laser high-harmonic generation and can be described using the Keldysh ionisation theory(*33*). Based on the Fowler-Nordheim (F-N) field-emission mechanism, nanotip electron emission shows no apparent correlation between the electron emission efficiency and photon energy, thereby easing the constraints on the selection of viable optical sources(*34*). The photoelectron spectrum and pulse duration can be further controlled using $\omega$-$2\omega$ field shaping(*35*) which also partly relaxes the requirement for CEP stabilisation(*36, 37*). While all this makes nanotips an attractive source of electron pulse, further advancements in electron collection and mapping are necessary to implement these emitters in nanoscopic measurements.

Potentially, the most straightforward and robust approach to generate short nanoscale electron pulses would be via nanodevices bearing asymmetric work function electrodes. Combining such structures with ultrashort pulses would enable the generation of sub-cycle electrons from intense optical fields across a broad spectral range, with arbitrary pulse shape and duration(*34, 38-40*). Given the significant reduction in the requirements for the optical field, including no need for few-cycle pulses with CEP stabilisation, asymmetric work function devices can be seamlessly integrated into most conventional experimental setups, simplifying ultrafast nanoscale electron probes for molecular scale spectroscopy. However, the scalable and reliable manufacturing of such nanodevices presents significant challenges; when based on electron beam lithography or scanning probe microscopy techniques, fabrication becomes expensive and low-throughput(*41*).

In this study, we employ asymmetric on-chip nanostructures to generate ultrafast tunnelling current pulses. These nanostructure devices, called photosensitive asymmetric nanogaps (PANs) are fabricated using adhesion lithography, a simple, high-yield, and cost-efficient coplanar nanogap fabrication technique(*42*). Electron pulses in PANs were generated upon the nanogap irradiation with different ultrafast lasers, without either CEP stabilisation or the need for external voltage bias. PANs performance, resemblant of a double Schottky diode with a sub-10-nm empty-gap tunnelling barrier, demonstrated rectification of optically induced current under short-pulse illumination. By tuning the wavelength from 266 to 6700 nm and pulse duration from <10 to >100 fs, we show that pulses of electrons are indeed ultrafast and produced by the Fowler-Nordheim field emission, rather than thermal expansion, thermoelectric effects, or multi-photon assisted tunnelling. We demonstrate the practical use of PANs for optical detection, tracking the ultrafast electronic dynamics of nanomaterial systems located in the nanogap, and suggest future applications of PANs technology.

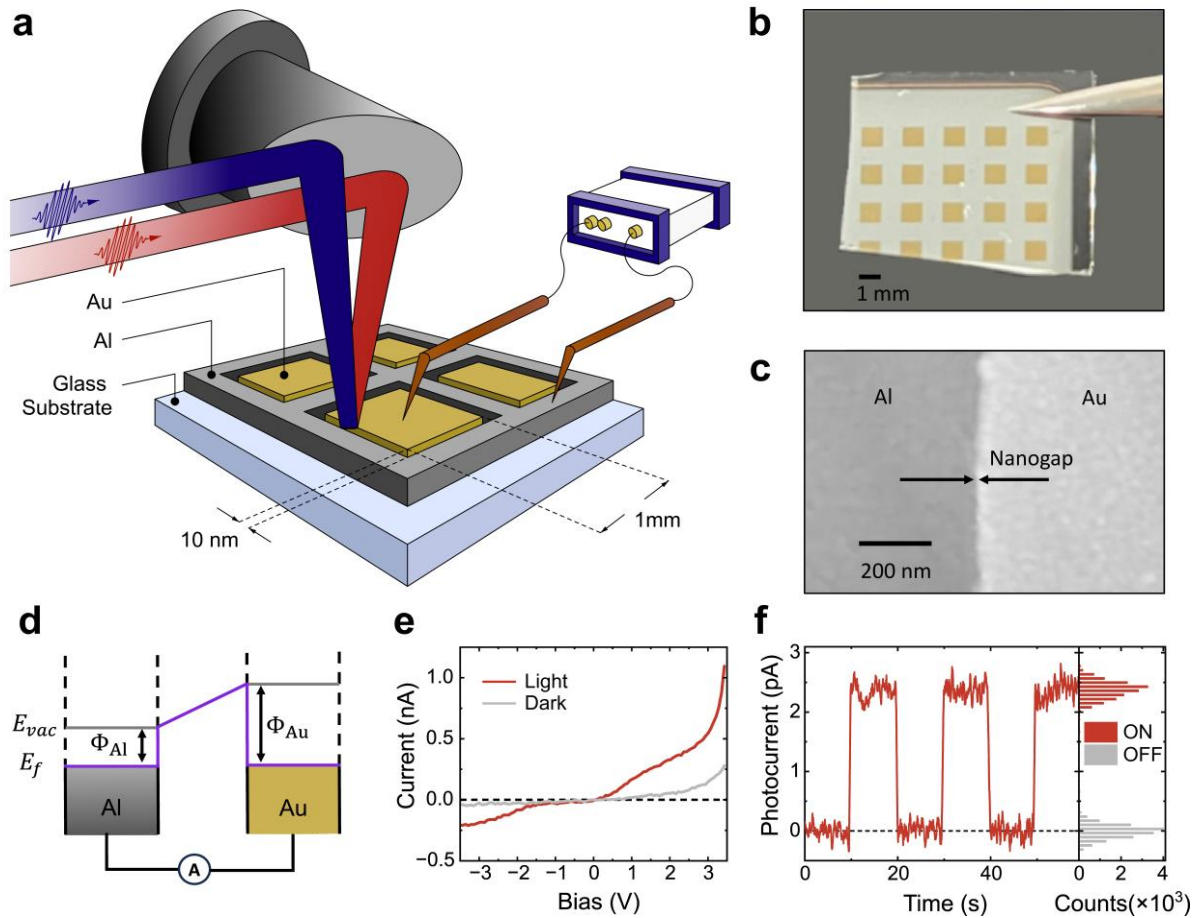

*Fig. 1 Strong-field tunneling in a PAN device: a) Schematic of an Au-nanogap-Al PAN device under ultrafast optical excitation. b) Photograph of multiple nanogap devices on one chip. c) SEM image of the nanogap device. d) Band diagram of nanogap diode comprising Al and Au metal electrodes. e) J-V characteristics of PAN diode with/without illumination. f) On-off test under 1030 nm 190fs 10kHz pulsed irradiation showing typical photocurrent measured at 0V bias and its dispersion with 30ms averaging.*

**Photosensitive Asymmetric Nanogaps (PANs)**

To achieve CEP-free and bias-free photoinduced tunnelling currents, we engineered(*43*) PANs, metal-nanogap-metal diodes with an asymmetric structure (Fig. 1a). A planar quasi-2D gold-insulator-aluminium structure is fabricated through adhesion lithography on a glass substrate, as depicted in Fig. 1a-c (see SI for fabrication details). This technique enables easy and high-throughput fabrication of arrays of asymmetric metal electrodes with large aspect ratios without requiring costly and time-consuming processes and equipment, like e-beam lithography(*42*). Asymmetric electrodes fabricated with adhesion lithography have been used in GHz radiofrequency diodes(*44*), fast photodetectors(*45*), organic light-emitting diodes(*46*), and memory devices, including ferroelectric tunnel junctions(*47*). However, the strong-field effects at these nanochannels have not been studied so far.

The PAN chips are composed of a glass substrate covered by a thin (40 nm) layer of aluminium. The Al film does not cover the substrate completely; 1-mm square 'islands' of gold (40 nm thickness) are periodically patterned on the substrate, each Au island separated from aluminium by a nanogap. (Fig. 1a,b) The length of each nanogap is equal to the island's perimeter (~4 mm) and the width of the gap is ~10 nm based on SEM (Fig. 1c) and previous work(*43*). During the experiment, only a part of the gap (typically 50 μm in length) was illuminated by ultrafast laser pulses.

The significant nominal offset (~1.2 eV) in work functions of the metal counterparts (Au vs Al) leads to strong current rectification properties of PANs due to the asymmetric and nonlinear response to the external electric field. This effect can be seen, for instance, in the *J-V* characteristics measured without laser irradiation from -3V to +3V (Fig. 1e). The minimal current flow in the reverse voltage regime as well as the exponential increase in current at ~1.2V in the forward direction are reminiscent of diode characteristics. These characteristics ensure the electron emission rates are different upon inverting the direction of the optical field (Fig. 1d), thus producing net photocurrent without the requirement of CEP and bias voltage.

Under continuous-wave laser illumination, we did not observe any photocurrent (see SI), indicating the device is not susceptible to heating-induced current modulation, which is one of the key effects limiting externally-biased optically-assisted STM approaches.(*48, 49*) Upon ultrafast laser irradiation, the *J-V* characteristics of the PAN change (Fig. 1e), showing a distinct photocurrent response in on-off tests (Fig. 1f). This suggests optical activation of tunnelling electrons emission via the strong electric field. We did not observe a discernible change in photocurrent as the ~50 μm beam spot was scanned along the nanogap.

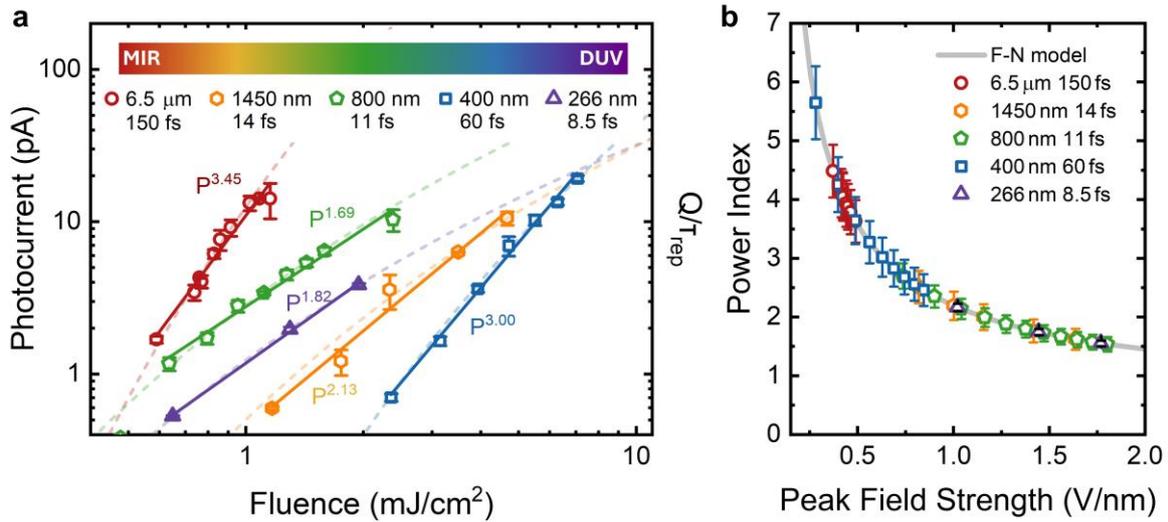

*Fig. 2 Optically induced currents in PAN devices: a) Ultrafast laser-induced photocurrents at zero external bias. Different colors/scatters correspond to experiments on various laser systems with different wavelengths of irradiating light (from UV to IR) and pulse durations. The solid lines are power-law fits to the data with the power index indicated. The dashed lines are the fits using the F-N field-emission model. b) Nonlinear power index and peak field strength calculated based on a) and field emission model, taking into account the wavelength, focal spot size, pulse energy, and pulse duration in different experiments. The solid line represents a global fit using the field-emission model for all experimental data collected across the optical spectrum.*

**Mechanism of photocurrent generation in PANs**

To explore the limits of PANs applicability and to investigate the mechanism behind the electron pulses, we have studied the efficiency of photocurrent generation depending on the photon energy, pulse energy, and pulse duration. We test the nanogap using various ultrafast laser systems (Supplementary Table S1), encompassing wavelengths ranging from 266 nm (deep-UV) to 6.5 μm (mid-IR), with pulse durations varying from ~8 fs to over 100 fs.

Fig. 2a presents the dependences of photocurrent on the radiant exposure for the different laser systems used in the absence of the external bias voltage. Across a very broad UV-to-IR spectral range, under ~0.2-20 mJ/cm² illumination, the ~0.3-30 pA level photocurrents are achieved across the nanogap. The curves are roughly linear in log-log plot and can be fitted by the power law function with the power indexes ranging between 1.7 and 3.4. The 2-fold variation in the power index is much smaller than 15-fold variation in the incident photon energy, indicating that the underlying mechanism is unlikely to be based on a multiphoton process. Different curve slopes observed for the lasers of similar wavelengths indicate that the light pulse intensity and pulse duration both affect the systems' behaviour. Our measurements indicate that the peak light intensities (and, correspondingly, the peak electric field in the nanogap) rather than photons' energies play the dominant role in the photocurrent generation. When the power index (the exponent of power dependence) is plotted as a function of the calculated peak electric field achieved by the optical pulse, the results from all laser systems comprise a single trend (Fig. 2b), which can be well reproduced using a field emission model discussed below.

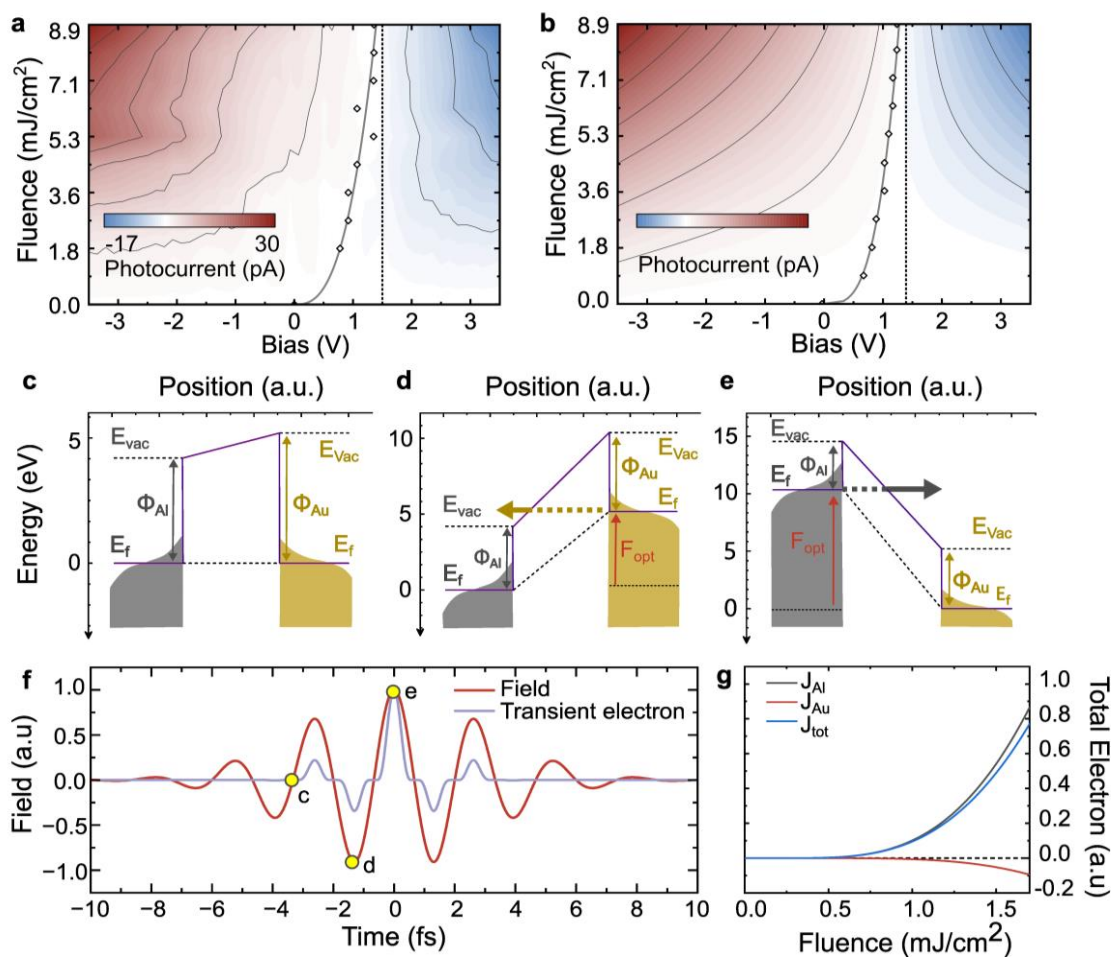

*Fig. 3 Field-emission model of photocurrent generation in PANs: a) The 2D map of the amplitude of photocurrent as a function of external bias, $V_b$, and illuminating fluence, $F$. The diamond line marks the minimum photocurrent position. b) Simulation of 2D map from a) based on photoemission model. c-e) Time snaps of band structure under ultrafast pulses (see f). The top of the grey-shaded area corresponds to the Fermi level of the aluminium electrode while the top of the gold area corresponds to the Fermi level of the gold electrode. f) Optical field and the electron flux as a function of time during the ultrafast irradiation of PAN with timings corresponding to c)-e) marked. g) The simulated power-photocurrent relationship, extracted by different electron emission rates from the two electrodes according to f).*

To identify the specific mechanism governing the generation of ultrafast photocurrents, we considered the signal dependence on the key experiment parameters, namely the light wavelength, external voltage, and peak field strength. The potential effects that may contribute to the photocurrent are: (i) slow modulation of the barrier due to thermal expansion and related phenomena; (ii) multiphoton-assisted tunnelling below the barrier energy; (iii) multiphoton-assisted electron emission with photon energy surpassing the barrier, or (iv) Fowler-Nordheim emission via modulation of the ionisation barrier by a strong optical field. The nonlinear response of photocurrent upon the illumination and its weak dependence on the illumination wavelength indicates that sample heating and heat-induced nanogap modulation have negligible contributions to the observed signals.

Photon-assisted tunnelling can be considered an alternative mechanism, which entails the absorption of one or more photons by electrons at the electrode interface, facilitating their tunnelling across the nanogap even when their energy is below the vacuum level on the opposing side of the tunnelling barrier. However, our calculations of photon-assisted tunnelling currents for pulsed optical fields based on Bessel solution through Tien-Gorden theory (see SI) contradict experimental data. The theory predicts a much higher order power law (power index of ~8) for photocurrent dependence on illumination, compared to the power indexes of 1.6-3.5 observed in the experiment (Fig. 2a). Another observation, suggesting that photon-assisted tunnelling is unlikely, comes from the high amplitude of the electrical current (1000 electron/fs/$\mu m^2$) during the optical pulse duration at relatively modest DC electric field across the nanogap. Such high currents at low voltage would not be possible if photoinduced electrons have energies below the vacuum level(*50, 51*).

Finally, we consider electron emission mechanisms via strong-field ionisation barrier modulation and multiphoton ionisation. These two mechanisms are well investigated for gases in strong electric fields, and they can be distinguished using the Keldysh parameter(*52*). By taking the work function of electrodes as an equivalent to ionisation energy of a gas, we find the Keldysh parameter is between 0.001 and 2.5 (Supplementary Fig. S9), suggesting that most of our experimental data lie within the strong-field region rather than the multiphoton region. Consequently, we conclude that the variations in the barrier induced by the optical field and the subsequent emission of electrons dominate the photocurrent generation.

Fig. 3 shows how F-N mechanism leads to the generation of a rectified current pulse in the oscillating field of an optical pulse and illustrates its predictive power. Fig. 3a shows the photocurrent generated in PAN by 800-nm 35-fs optical pulses as a function of irradiation power and external bias. Notably, the photocurrent is present even at zero bias, with its amplitude and direction of controlled by applied voltage. The photocurrent dependence on irradiating flux still follows a power law, but increased bias gradually diminishes the value of the power index (see SI), until the current becomes negligible and then changes sign at the open-circuit voltage. All the observed trends can be reproduced (Fig. 3b) with a F-N field-emission model described in detail below.

Without external field (Fig. 3c), electron populations within both electrodes are in equilibrium and no current flows. The laser pulse irradiating the nanojunction effectively applies rapidly oscillating, symmetric positive/negative bias voltage. Under a positive voltage (Fig. 3e), the laser-induced voltage elevates $E_{f\,Al}$, causing electrons in Al with energies higher than $E_{vac\,Au}$ to tunnel through the barrier and undergo field emission. In our schematic, the optical field bias magnitude is high enough that the tunnelling region (E<$E_{f_{Al}}$, $E < E_{vac_{Au}}$) does not contribute significantly to the overall process. Under negative voltage conditions (Fig. 3d), the gold electrode also emits photoelectrons; however, due to the higher barrier, the resulting current is significantly smaller than in the positive voltage case (see SI for full detail). Due to the junction asymmetry, even a fully symmetric pulse can generate photocurrent at bias-free conditions. This current can be modulated by the control of CEP, but photocurrent generation in the PAN does not rely on CEP stabilisation. The FN model shows perfect agreement with measured data as demonstrated in Fig. 2a,b and 3a,b confirming our interpretation in full.

**Broadband ultrafast pulse autocorrelation measurements with PANs**

The transfer of field-emission electron pulses across ~10-nm gap is expected to occur at sub-cycle timescales (see Fig. S12 in SI) making the electron pulses very short and thus attractive ultrafast probe for both irradiating optical fields and intra-gap environment. We verify the fast response time of PANs using optical autocorrelation, as the process is too fast to be measured using electronic detection.

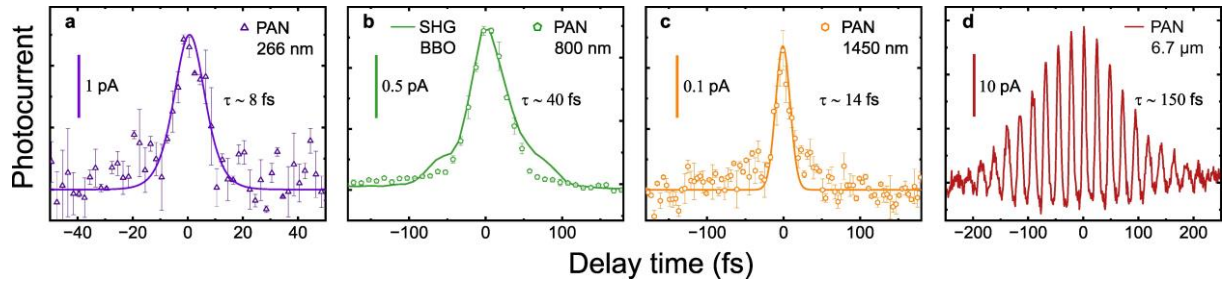

*Fig. 4 Characterisation of ultrafast optical pulses using the same PAN device. Autocorrelation signals for: **a)** deep UV ~8 fs 266nm pulse, measurements were done in vacuum at $10^{-6}$ bar; **b)** red ~40 fs 800 nm pulse; solid line is the autocorrelation measured using second harmonic generation in BBO crystal; **c)** NIR ~14 fs 1450nm pulse. **d)** Interferometric measurement of ~150 fs 6.7µm pulses. All the tests are under no electrical bias.*

Fig. 4 presents baseline-corrected two-pulse intensity autocorrelation under non-collinear geometry (a-c) and the interferometric autocorrelation (d) under collinear geometry for various laser systems. All the autocorrelation traces demonstrate symmetric shapes and no long-lived component in the signals even for 8-fs pulses, indicating the fast response and absence of an electronic population built-up or heating effects. For 800-nm pulses, we compared photocurrent autocorrelation to the BBO second harmonic generation (SHG) autocorrelation measurement and, after the correction for nonlinear order, observed a good agreement between the data (Fig. 4b). This was achieved without any need to satisfy phase-matching conditions in PAN, highlighting the possible applicability of tunnelling-based detection for characterisation of short and broadband pulses in IR and deep UV range where the use of non-linear crystals can be problematic (Fig. 4a, c). To demonstrate the robustness of the approach, we systematically varied the pre-chirp in the amplifier and successfully monitored the alteration in the shape of the optical pulse (Fig. S5). We also show the use of the same device to detect interferometric response at a wavelength of 6.7 µm (Fig. 4d). Its Fourier transformation can be used to retrieve the spectrum, which agrees well with the results obtained using a HgCdTe photodiode (see SI). The above pulse characterisation involves modest laser powers, a low level of non-linearity, and no need for CEP stabilisation or sub-10fs compression, which makes PANs more versatile and accessible than existing state-of-the-art pulse sampling techniques(*53*).

**Time-resolved nanoscopy of semiconductor films and quantum dots in PANs**

The possibility of generating ultrafast nanoscale photocurrent pulses in PANs carries a clear potential for nanoscale spectroscopy applications. The generated electrons can be used to probe the electronic and structural dynamics of molecular-scale systems entrapped in the nanogap. The fact that electrons can be generated by light of practically any wavelength makes this approach highly flexible and versatile. The optical-pump electron-probe spectroscopy, has been previously demonstrated for STM-based systems.(*19*) We extend this approach to monitor charge dynamics in nanosystems within the PAN's 10-nm channel.

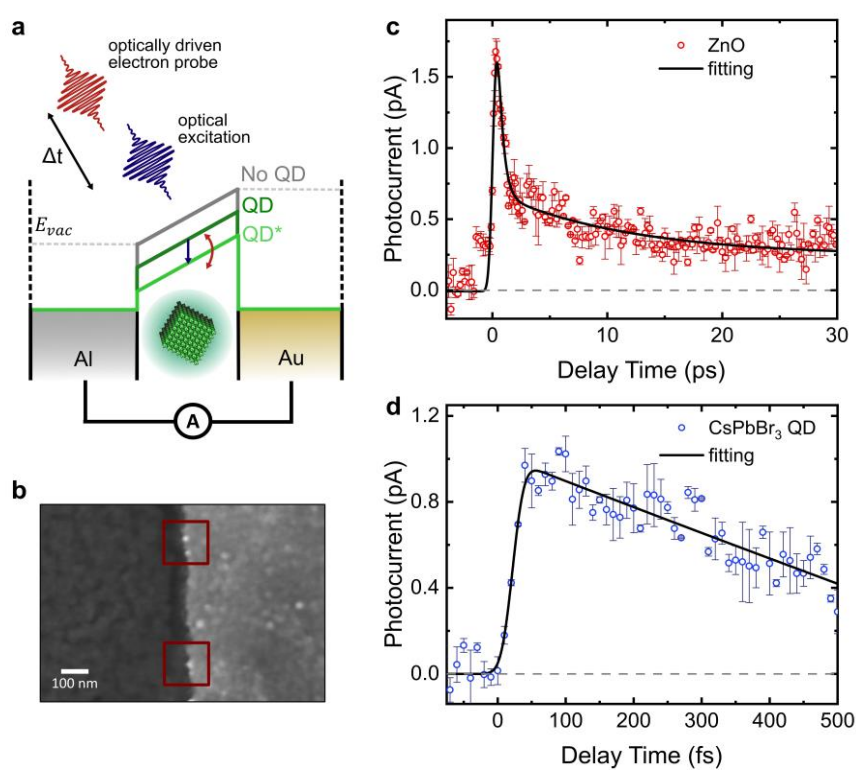

*Fig. 5 Time-resolved measurements of electronic dynamics inside the nanogap using photoelectrons. a) The conceptual presentation of optical pump – electron probe nanoscopy. b) SEM image of CsPbBr$_3$ QDs in the PAN. Red squares highlight the individual QDs located in the nanogap area. c) 400nm optical pump, 6.7um tunneling electron probe measurement of electronic dynamics of ZnO. d) 400mm pump-800 nm tunneling electron probe measurement of sub-picosecond conductance dynamics of CsPbBr$_3$ QD.*

Fig. 5 summarises the proof-of-principle tests of our nanoscopy approach for small (~10nm) CsPbBr$_3$ quantum dots (QDs) and polycrystalline ZnO films. Using a simple spin-coating technique, we directly deposited ZnO polycrystalline film or individual QDs inside the nanogap (Fig. 5a,b). When CsPbBr$_3$ or ZnO are located inside the gap, the dark tunnelling current in the PAN changes due to the changes in the tunnelling barrier, consequently leading to alterations in the *J-V* curves (see SI). The photocurrent also changes and can be further modulated by bringing the material to the excited state using pump pulse, as this leads to further modulation of the tunnelling barrier. During the pump-probe experiment, a weak high-photon energy pump pulse photoexcites material in the nanogap without inducing tunnelling. Then, an intense but low-photon energy probe pulse generates tunnelling electrons to monitor the changes in nanomaterial excited states without its direct electronic excitation.

For ZnO, we used 400 nm pump pulses for above-gap excitation of the material and 6.7 µm below-gap probe. As ZnO is transparent in mid-IR, such probe did not interact with ZnO directly, but generated electron pulses in the PAN. As the concentration of the excited states promptly goes up after photoexcitation and then decays on multiple timescales (from 200 fs to 10s of ps) due to recombination, we observed the change in tunnelling current on timescales (Fig. 5c), in agreement with earlier studies(*43, 54*).

To emphasise the potential of PANs to characterise individual nanosystems, we deposited $CsPbBr_3$ perovskite QDs inside the nanogap and addressed their electronic dynamics. Based on SEM measurements we expect the density of QD along the slit to be in the order of 3 QD per 1µm, indicating that just ~100 QDs contributed to the signal in each experiment. We employed ~10-fs 400 nm pulses as excitation and similarly short 700-900 nm pulses for generating tunnelling-electron probe. Upon excitation, we observed that the amplitude of photocurrent instantly increases by ~1 pA and then decays with a roughly 300 fs time constant (Fig. 5d). We assume this is caused by the excitation of electrons in the quantum dot conduction band through 380 nm illumination. The observed lifetime was consistent across various samples and was notably shorter than the carrier lifetimes reported based on transient absorption measurements.(*55*) These results, however, align qualitatively with the dynamics of carrier cooling in perovskite nanomaterials(*55*) and with photoconductivity measurements on perovskite thin films obtained via transient terahertz spectroscopy(*56*). We, therefore, attribute the observed signal to the formation of hot carriers, followed by their cooling and/or extraction(*57-60*).

**Conclusion**

The presented results show that combining on-chip asymmetric nanogap structures, PANs, with ultrafast illumination provides a versatile method to generate ultrafast pulses of tunnelling electrons via the broadband, Fowler-Nordheim photoelectron emission. PANs are fabricated with an inexpensive, facile, high-throughput technique, called adhesion lithography, and operate without the need for CEP stabilisation or an external electric bias, making them applicable in any ultrafast laser lab. We have also demonstrated the practical use of PANs for broadband characterisation of optical pulses and ultrafast electron nanoscopy, using perovskite QDs and ZnO as model systems. The developed approach also opens new research possibilities in photonics, including nonlinear-crystal-free optical pulse characterisation and nanoscale imaging. We anticipate that PANs will emerge as a cost-effective, durable, and adaptable foundation for integrated ultrafast broadband optoelectronics and for ultrafast spectroscopy on individual nanosystems.

**Acknowledgements:**

We thank Maxim Pchenitchnikov and Ernest Pastor for valuable discussions. AAB acknowledges support from the Royal Society via the University Research Fellowship and Leverhulme Trust via the Philip Leverhulme Prize award. This project received funding from the European Research Council (ERC) under the European Union's Horizon 2020 research and innovation program (Grant Agreement 639750/VIBCONTROL) and from UKRI/EPSRC (ActionSpec, grant ref: EP/X030822/1). The project was also supported by the EPSRC New Horizons Grant (TRON, grant ref: EP/V049070/1). Dmitry R. Maslennikov acknowledges funding by the Imperial College London President's PhD Scholarships. CF and JPM acknowledge funding support from UKRI/EPRSC EP/V026690/1, EP/T006943/1, and EP/R019509/1. DGG and EM acknowledge support from the UKRI Future Leaders Fellowship Grant (MR/V024442/1).

**Supplementary Materials**

# S1. Fabrication of Asymmetric Al/Au Nanogap

**Asymmetric Al/Au nanogap fabrication:**

The Al-Au asymmetric nanogap is fabricated using adhesion lithography. Initially, a 40nm thick Al electrode is thermally evaporated onto a glass substrate. Photolithography, followed by wet etching, is then employed to establish the initial pattern. Subsequently, the substrate is immersed in a 1-mM solution of the self-assembled monolayer (SAM) octadecylphosphonic acid in 2-propanol to generate a hydrophobic layer on the aluminium surface. A 5nm Al layer followed by a 35 nm thick layer of Au is subsequently deposited. Following the deposition of gold, an adhesive film (from Photonic Cleaning Technologies) is applied to the wafer, left to dry, and later peeled off manually, also removing the Al/Au part that is in contact with the SAM-functionalised Aluminium. Finally, the organic SAM layer can be removed through UV-ozone treatment, resulting in a 10-20nm nanogap.

## S2. Reproducibility of nanogap size on PAN

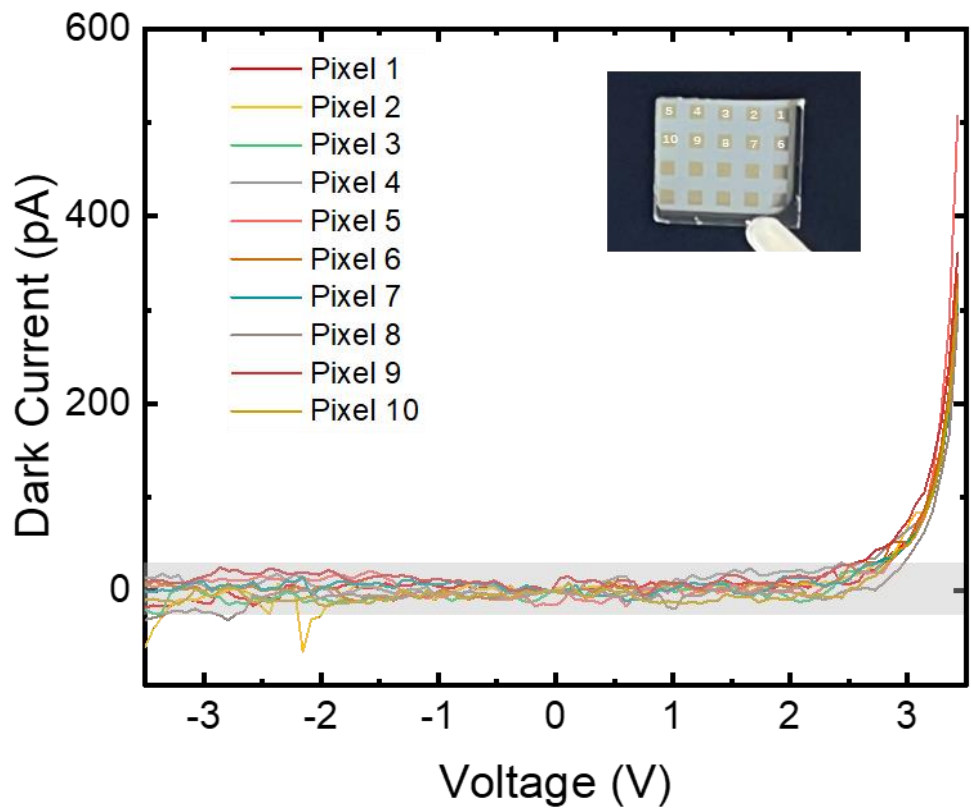

Fig. S1 I-V characteristic of 10 distinct pixels of the same chip. All tests are under no illumination. The turning point (~3V) shows gap size variation is controlled. The shaded current range is within the instrumental limitation of the amplifier under 0Hz.

The dark current level on the order of 0.2 nA at 3 V allows lock-in detection of optically induced currents with accuracy below 1pA. Notably, the bias DC current contribution stemmed from the entire PAN diode with a total working area of 4000 μm (Au perimeter) x 40 nm (thickness). In comparison, conventional tunnel junctions with smaller effective areas possess typical widths in the ~nm range, resulting in current levels of 1nA/0.1V. This difference in the level of current rules out the presence of any <2 nm spaced hotspots in PANs.

*J-V* measurements from multiple PANs confirm the consistency of photocurrent distribution along the nanogap. To validate the size reproducibility of nanogap size on PAN, we tested and compared the gap size variation on 10 distinct devices. The slope and current level could indicate the homogeneity of PAN due to the distance-field strength relationship on Fowler-Nordheim (F-N) tunneling. Therefore, we perform the J-V sweeps of different PAN samples within the voltage range of -3.5V to 3.5V. Through analysis of the consistency in current magnitude, we conclude that the size repeatability of the nanostructures is ensured.

## S3. CW illumination control experiment.

As a control experiment, we irradiated samples with a 670 nm CW laser of average power 2 mW using a lens of 10 mm focus. The CW laser was modulated at 1 kHz through an optical chopper. Due to the absence of voltage bias requirements and the relatively small length variation dZ/Z resulting from larger gaps (~10 nm), no observable thermal flow was generated by the CW laser (Fig. S2). In contrast, plasmonic heating induces thermal expansion on nm-spaced nanowires or nano junctions formed between the STM tip and sample plane, which is more prominent.

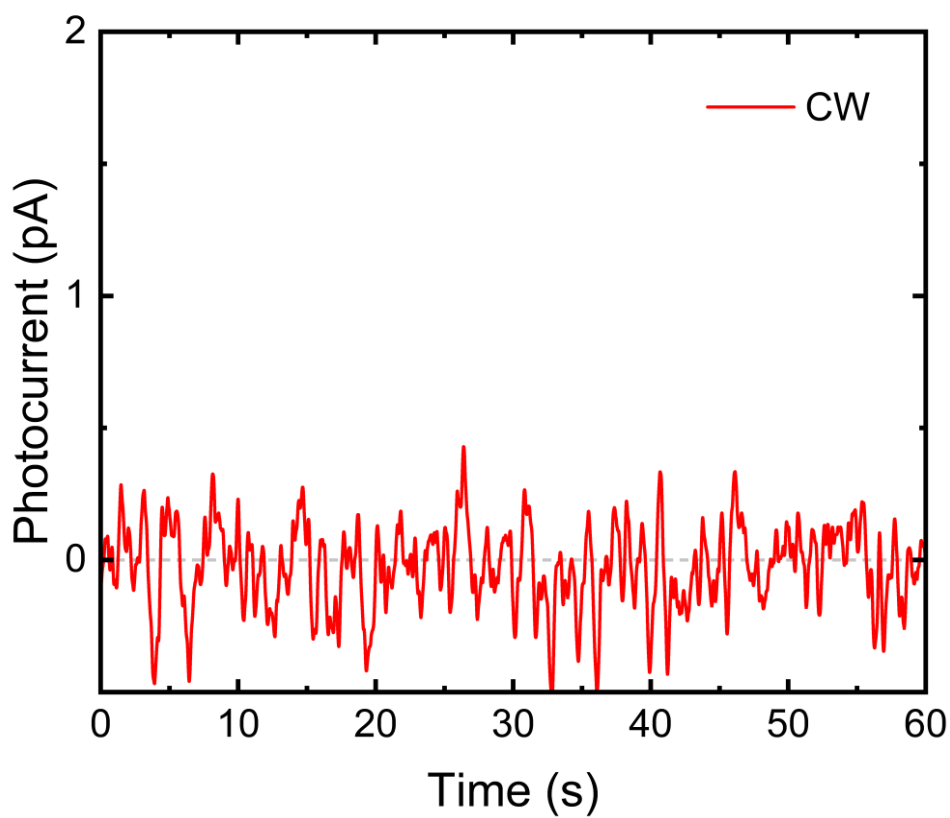

Fig. S2 Photocurrent test under CW illumination. On-off test under 670nm CW laser, chopped at 1kHz.

# S4. Numerical simulation of possible optical-pulse-induced thermal expansion effects with/without bias.

When the focused laser pulse irradiates the metallic nanostructures, the electron-phonon coupling will ultimately induce lattice heating, leading to changes in nano-size dimensions that may impact the level of electrical current. Here, we illustrate that, under unbiased conditions, the changes in the signal due to thermal expansion can be neglected, as depicted in Fig. S3.

When no DC bias is applied (Fig. S3a), if we consider thermal expansion as a signal on the microsecond scale, its temporal product with the original voltage signal will influence the final shape of the electrical pulse. However, the photo-voltage, being the sole voltage signal, does not contribute any signal beyond the duration of the optical pulse.

In the presence of bias (Fig. S3b), the direct current (DC) signal, treated as a continuously existing signal, continuously convolves with the thermal expansion signal in the time domain. Assuming a constant length of thermal expansion at the same power, as the nanostructure dimensions decrease, the impact of this signal becomes more pronounced. This slow signal may also result in the temporal resolution of electronic pulses under bias being influenced not only by the optical pulse itself but also by the continuous presence of the DC signal.

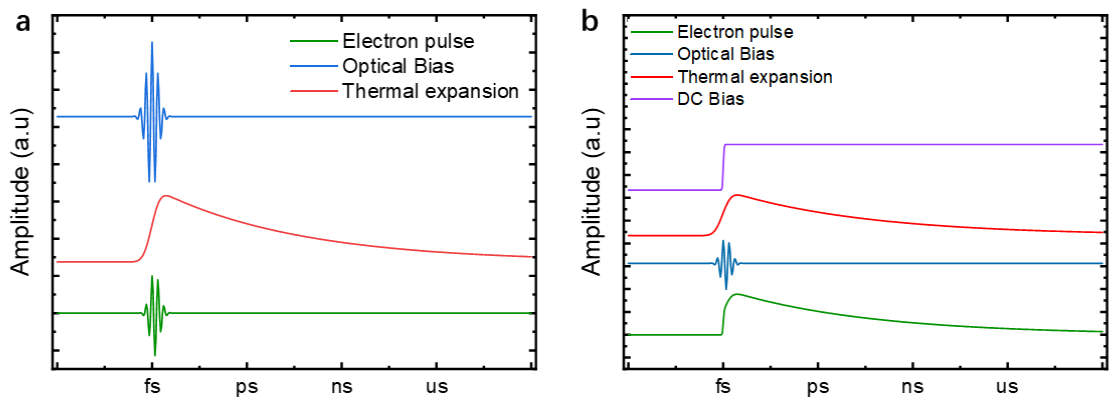

Fig. S3 Electron pulse duration (green) when a slow thermal expansion is present. a) When no bias is applied, the tail part of the expansion will not induce any slow current b) When bias (purple) is applied, the ns to $\mu$s tail will generate a slow and strong electron pulse.

# S5. Representative experimental setup. Laser systems used.

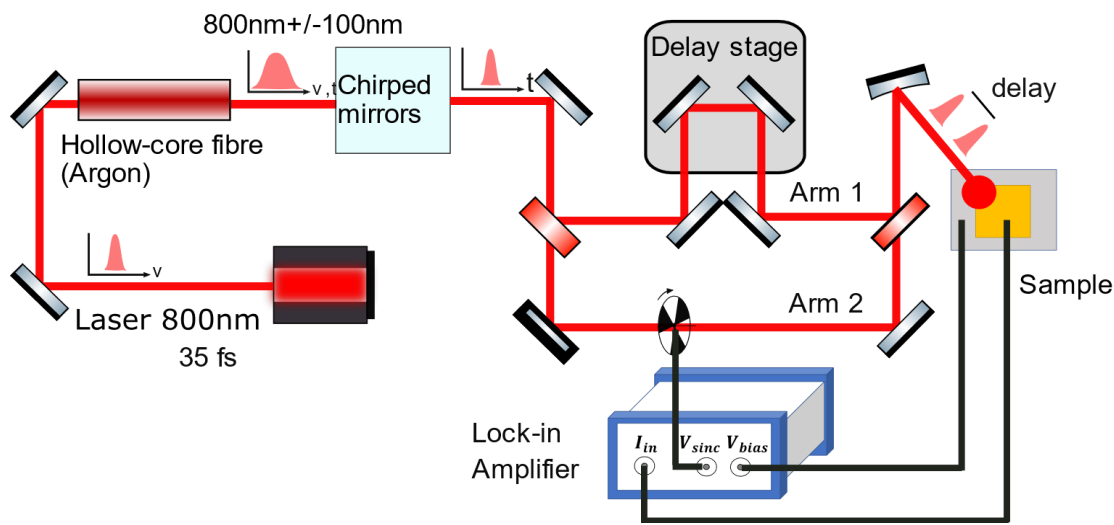

Fig. S4 Schematic representation of the Autocorrelation setup with photocurrent detection. The laser system here is Astrella (Coherent) 4kHz and the pulse compression is done via hollow-core fiber and a pair of chirped mirrors (800nm 10fs).

| Wavelength | Pulse duration | Generation |
|---|---|---|
| 266nm | ~8.5 fs | Red Dragon (KM labs) at 1kHz + Optical Parametric Amplifier + Hollow Core Fibers |
| 400nm | ~100 fs | Astrella (Coherent) at 4kHz + BBO crystal for SHG |
| 800nm | ~35 fs | Astrella (Coherent) at 4kHz Fundamental pulse |
| 800nm few cycle | ~10 fs | Astrella (Coherent) at 4kHz +Hollow Core Fiber |
| 1030nm | ~200 fs | Pharos (Light conversion) at 200 kHz |
| 1.45um | ~10 fs | Red Dragon (KM labs) at 1kHz + Optical Parametric Amplifier + Hollow Core Fiber |
| 6um | ~100 fs | Astrella (Coherent) at 4kHz + TOPAS OPA + $AgGaS_2$ for DFG |

*Table S1. Wavelength, pulse duration and generation methods under working systems.*

We carried out photocurrent measurements using different laser sources to achieve broadband coverage of the spectrum spanning from DUV 266nm up to MIR 6.7μm. A hollow core fiber filled with Argon is used to generate ultrafast pulses. The pulse duration and the laser system used for each wavelength are shown in Table S1 above. For example, Fig. S4 shows the photocurrent measurement performed for the case of 800nm with 10fs pulse duration. For this case, Astrella (Coherent) at 4kHz is used along with hollow-core fiber pumped differentially with Argon and a pair of chirped mirrors. For all wavelength delay between 2 pulses is controlled by translation stage. The device is mounted in a commercial two-electrode probe stage (Linkham Scientific); the metallic stage can also work as a Faraday cage, which cancels external electric field interference. Laser pulses irradiated the nanogap chip. For current and photocurrent recording, we use a 500 kHz digital lock-in amplifier (MFLI, Zurich Instruments) with built-in auxiliary bias output. The demodulator frequency is selected to 0Hz for DC current recording and selected for external synchronization (1kHz or 4kHz) for photocurrent recording. For both current and photocurrent measurement, we choose 10MV/A for the transimpedance gain, which gives a noise level of 150 $fA/\sqrt{Hz}$ at 1kHz.

## S6. I-V and photocurrent characteristics modelling using two-electrode Fowler-Nordheim emission theory.

In Fowler-Nordheim (F-N) tunneling, electrons overcome ionization potential under a strong electric field, tunnel through a thinned barrier and accelerate through an insulating medium. F-N tunneling requires an extremely strong electric field (on the order of $10^8$ V/m, requiring laser intensities > $10^9$ Wcm$^2$), which corresponds precisely to the field strength across nanogap in our experiments. Considering the configuration of our double Schottky barrier diode devices, we describe the electron emission process in two F-N tunneling terms as shown below. The first term corresponds to electron emission from Aluminum to gold and the second term corresponds to electron emission from gold to Aluminum. For the case of external bias being applied, this means that first term and second terms are assigned to positive and negative bias. For the case of the laser-induced field, the two terms are assigned to the opposite directions of the electric field, the positive and negative part of field cycles.

$$j(F) \propto \Theta(-F)|F|^2 \exp\left(-\frac{4\sqrt{2m_e}\Phi_1^{\frac{3}{2}}}{3\hbar e\beta|F|}\right) - \Theta(F(t))|F(t)|^2 \exp\left(-\frac{4\sqrt{2m_e}\Phi_2^{\frac{3}{2}}}{3\hbar e\beta|F|}\right) \quad (S1)$$

where $e$ is the elementary charge, $m_e$ is the electron mass, $j$ is the instantaneous current density crossing the nanoscale junction, $\beta$ is the enhancement factor, F is the field strength, $\Phi_1$ (4.12eV) is the work function of Al and $\Phi_2$ (5.2eV) is the work function of Au. $\Theta$ is the Heaviside function and is used here to differentiate between negative and positive bias or the opposite directions of laser electric field during its cycles. Therefore, $\Theta(-F) = 1$ and $\Theta(F(t)) = 0$, when bias or laser electric field has direction from Al to Au, and $\Theta(-F) = 0$ and $\Theta(F(t)) = 1$, when bias or laser electric field has direction from Au to Al. In the case of biased (no optical field) experiments, the effective area of the current is proportional to the entire 4mmx40nm nanochannel. In experiments with optical fields, the photocurrent only originates from the illuminated spot area. For a single pair of Al-Au emitters, the current term from Al to Au is determined by work function $\Phi_1$, and the current term from Au to Al is determined by the work function $\Phi_2$. For barriers $\Phi_1$ and $\Phi_2$, to simplify the model, we only consider the dominant processes and we do not consider the spatial variation of the barriers, such as bending at the interface.

The field strength F is the sum of the optical field and the DC electric field,

$$F = E_{opt} + \frac{V_{DC}}{d} \qquad (S2)$$

where $E_{opt}$ is the laser electric field, $d$ is the nanogap distance, and $V_{DC}$ is the DC biased. The optical field, $E_{opt}$, is kept independent of the distance for simplification. For the case of no light,

$$F = \frac{V_{DC}}{d} \qquad (S3)$$

For the I-V curve fitting (no light), it is necessary to include the enhancement factor $\beta$ between 10-100. $\beta$ varies with changes in emitter material and geometry (18) and cathode-anode distances. Therefore, a fixed factor from I-V fitting for different devices is extracted. In the optical measurement analysis shown below, a fixed $\beta$ value is used for all wavelength fitting. We did not induce a wavelength dependence on $\beta$ due to plasmonic enhancement for optical measurement, because Aluminium plasmonic resonance is supposed to be strong in DUV and it will not be a dominant varying parameter within a wider spectrum span.

However, if the impact of large gaps on different kinetic energy electrons, such as the Simple man's model (9), or effects like plasmonic enhancement, needs correction, the distance will also affect the optical field. The field strength of the DC electric field is directly related to the bias voltage $V_{Dc}$ and gap length d. The simulated and experimental I-V characteristic of the nanogap without applying an optical field is shown in Fig. S5.

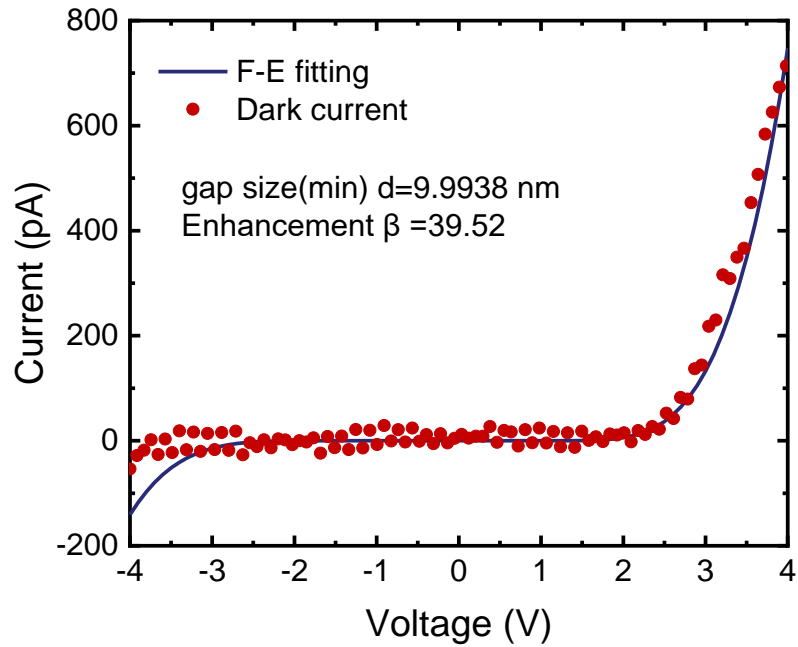

Fig. S5 Measured dark current (red dot) and simulated J-V characteristic (dark blue line) in Fowler-Nordheim regime. They show good agreement when only DC bias is applied.

## S7. Fowler-Nordheim modeling of PANs photocurrent under different wavelengths irradiation [Fig. 2 of the main text]

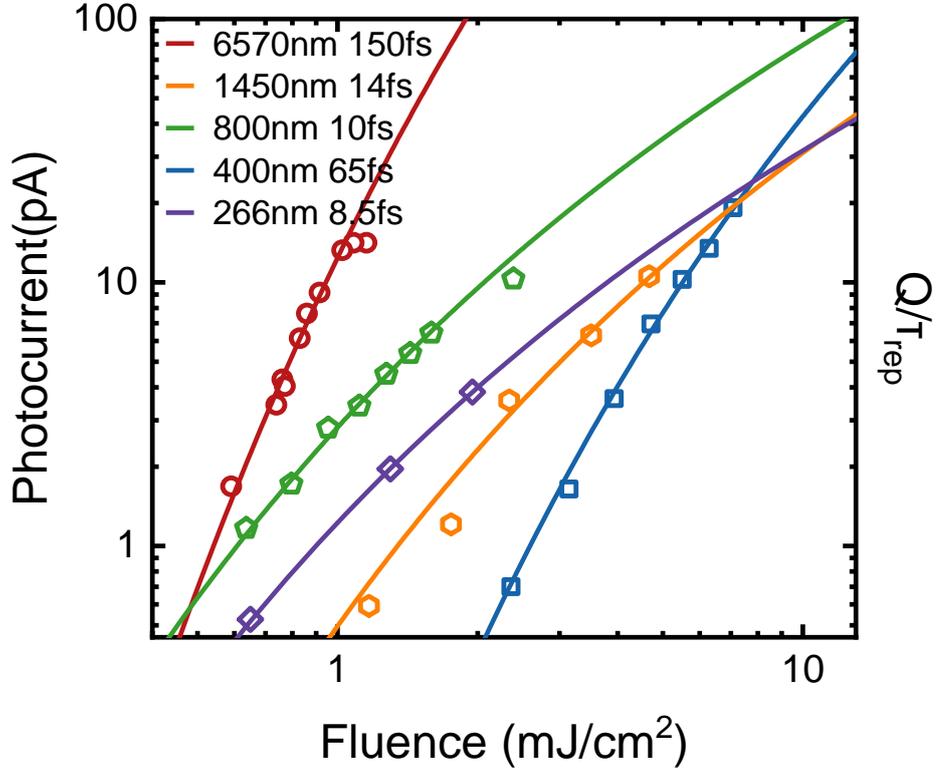

Fig. S6 Model (lines) compared to measured experimental data for photocurrent (points) for different wavelengths and fluences. (Main text Fig 2.1).

When laser pulse is present, and there is no applied DC biased then the field strength is,

$$F = E_{opt} \qquad (S4)$$

But laser electric field is a function of time and space,

$$F(r,t) = E_{opt}(r,t) \qquad (S5)$$

Thus, the F-N expression becomes,

$$j(r,t) \propto \Theta(-F)|F|^2 \exp\left(-\frac{4\sqrt{2m_e}\Phi_1^{\frac{3}{2}}}{3\hbar e\beta|F|}\right) - \Theta(F(t))|F(t)|^2 \exp\left(-\frac{4\sqrt{2m_e}\Phi_2^{\frac{3}{2}}}{3\hbar e\beta|F|}\right) \qquad (S6)$$

Then, the average photocurrent readout J from the measurement experiment is:

$$J = \frac{Q}{\tau_{rep}} = A_{eff} \int_{R_{focus}} \int_{\tau_{pulse}} j(F(r,t)) dr dt \qquad (S7)$$

where $Q$ is the charge readout by lock-in amplifier, $\tau_{rep}$ is inverse of laser repetition rate, $R_{focus}$ is the measured beam waist on focal position and $\tau_{pulse}$ is pulse duration. $A_{eff}$ is a fitted pre-factor to match the amplitude of measured photocurrent, it is positively correlated effective number of emitters and it further varies at different mediums (e.g. vacuum to air) - more detailed in table S2. $j(r,t)$ is the instant current flow scale with the probability determined by F-E emission. $F(r,t)$ is the instant field strength and it is calculated as follows:

$$F(r,t) = E_{opt} + \frac{V_{dc}}{d} = \sqrt{\frac{2I_{opt}(r,t)}{c\epsilon_0}} + V_{dc} \qquad (S8)$$

$$I_{opt}(r,t) = I_{center} \exp\left(-4\ln 2\left(\frac{t^2}{\tau_{FWHM}^2} + \frac{r^2}{(2R_{focus})^2}\right)\right) \exp(i\omega t + \phi) \qquad (S9)$$

$$I_{center} = \frac{2P_{avg}}{f_{Rep} * 0.94\tau_{pulse} * \pi R_{focus}^2} \qquad (S10)$$

When $V_{dc} = 0$, $F(r,t)$ scale with intensity on position $r$ and at instant time $t$ (we put beam center as $r = 0$, and temporal peak position as $t = 0$). Here the $\tau_{FWHM}$, pulse duration FHWM, was taken from autocorrelation measurements (Fig. 4, Fig. S14, Fig. S15, Fig. S16). $P_{avg}$ is the measured average power. F(r,t) rapidly oscillates on every optical-cycle with a frequency $\omega$, but current direction is secured by the device asymmetry and the fact that we are working with multi-cycle pulses CEP will only have an influence on sub-cycle part of the pulse.

The fitting process is as follows: The F-E emission characteristic for PAN is determined by fitting DC measurement to j(F), as shown in the previous section. It is worth noting that we didn't use a varied system ($\beta$ and $d$) for different wavelengths. We mainly fit A to match the measured current to the model, but at the same time, we give a tolerable range to $R_{focus}$ considering radius changing due to the z position of the sample and the mismatch from ellipticity, which is detailed in Table S2.

The parameter A remained consistent within one order of magnitude except for the 6570nm measurement. For the MIR pulse dry air is used during the experiment to minimize MIR absorption by water molecules. This change of environment is suggested to cause this one-order difference, this is also reflected in the photocurrent as a function of fluence graph, where MIR pulse achieves high photocurrents in relatively low fluences (low field strengths).

| Central wavelength | $A_{eff}$ | Radius (um) |
| --- | --- | --- |
| 266nm | 3.82e6 | 57.53 (~50/100 on x/y axis) |
| 400nm | 9.70e6 | 86.72 (~50) |
| 800nm | 1.61e6 | 141.68 (~100) |
| 800nm + HCF | 1.64e6 | 124.12(~100) |
| 1450nm | 5.02e6 | 138.95(~175) |
| 6570nm | 8.71e7 | 83.64 (~100) |

*Table S2 Fitted $A_{eff}$ and Radius for all wavelengths. The bracket value is from the estimation of beam diameter and focus.*

# S8. Alternative mechanism check: Photon-assisted tunneling/emission (PAT/PAE)

Photon-assisted tunneling was discovered by Tien-Gorden and Tucker. This theoretical framework describes the relationship between the rectified current and the original DC current in the quantum picture of the AC field, as well as how variations in photon energy and light field intensity can influence the rectified current. As detailed by the following formula:

$$I(\omega) \sim \frac{e}{\hbar} \Sigma \left| J_n \left( \frac{V_{ac}}{n\hbar\omega} \right) \right|^2 I(n\hbar\omega), \quad (S11)$$

where $\hbar\omega$ is the photon energy, e is electron mass, Jn is the Bessel function of nth kind. Vac is the field strength.

First, we check the scenario of $n\hbar\omega < \varphi$. we observed that the rectified photocurrent decreases as field strength increases (Fig. S1). This results from the overall Bessel function value ($J_n$) for all orders, which decreases as the field strength increases. This is shown in Fig. S7b and c for the cases of low and high field strength, respectively. For a photon energy of 0.2 eV (MIR), we limited the photon order to 20 to set the Vac below 4 eV, which is still below the work function value of Aluminum. Therefore, we can exclude any PAT mechanism below the barrier as we experimentally observe that the rectified photocurrent increases within a field strength range of V/nm.

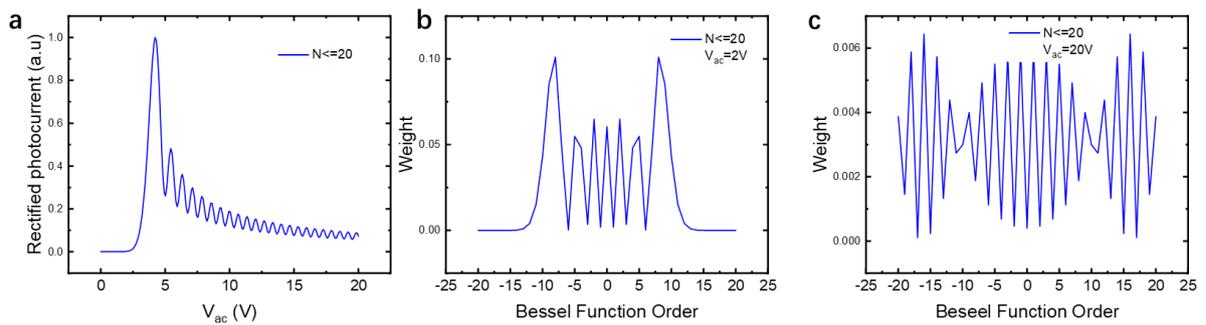

Fig. S7 Photon assisted tunneling with photon-number limitation (below barrier). a) Photocurrent variation with field strength b) Weight of contribution from different photon orders.

Second, we examine the regime of $n\hbar\omega > \varphi$ where PAE is realized. As shown on Fig. S8a there is an increase in photocurrent as field strength increases. This is a result of the higher overall Bessel function value (Jn) as higher orders are activated.

We now plot the total photocurrent as a function of the field strength for both regimes (PAT plus PAE), shown in Fig. S9. As the graph shows, the slope change is more dramatic for lower energy photons, and as we reach high peak field strengths, the slope becomes independent of the energy photon. None of these are observed in our case. Notably, for the 0.2eV photon (mid-IR pulse) Fig. S9 illustrates a power order of ~8 for the field strengths utilized in the experiment, significantly exceeding the order which has been experimentally measured, as shown in Fig. 2 and Fig. S6.

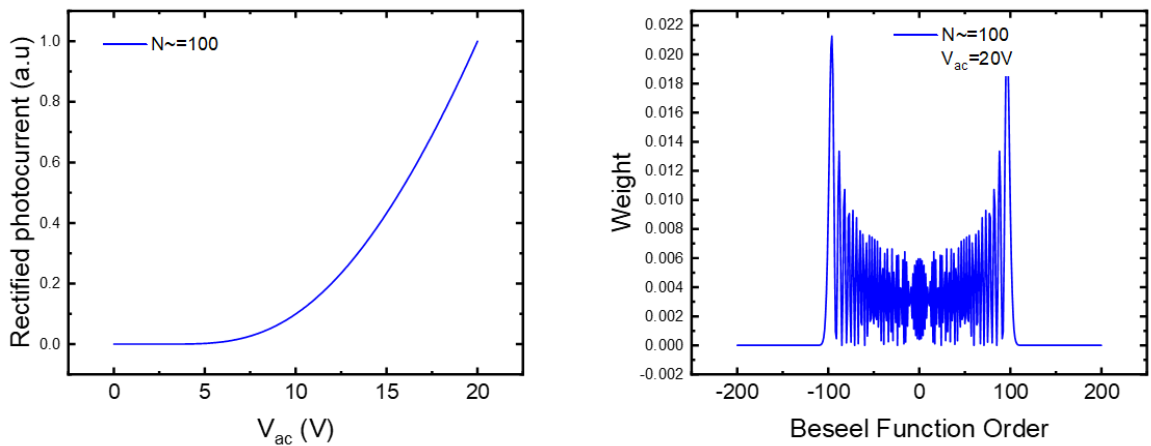

Fig. S8 Photon-assisted tunneling with no photon-order limitation (above barrier). a) Photocurrent variation with field strength b) Weight of contribution from different photon orders.

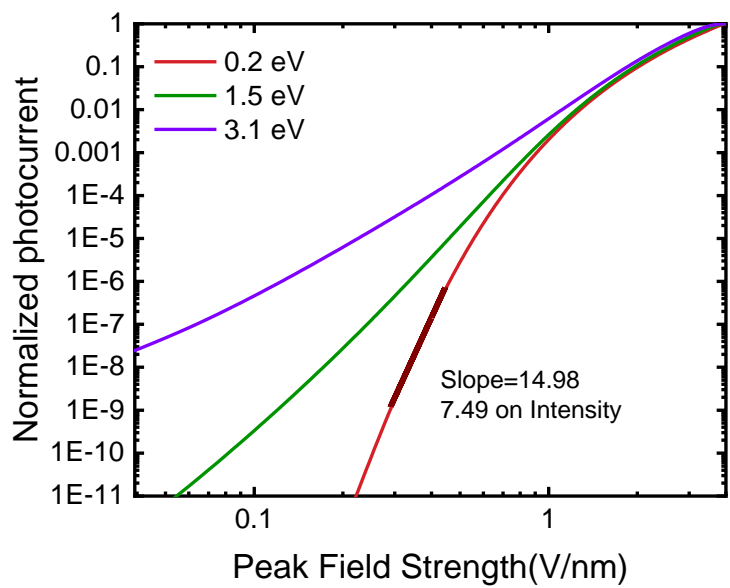

Fig. S9 Simulation of Photon-Assisted Tunneling yield for various photon energies. The dark red line covers the field strengths used in our experiments. PAT predicts a higher slope than the one experimentally measured (see Fig. S6 for the mid-IR pulse) and simulated by the F-E model.

## S9. Keldysh Parameter Estimation

The Keldysh parameter is defined as a function of ionization potential $I_p$ over ponderomotive kinetic energy $U_p$.

$$\gamma = \sqrt{\frac{I_p}{2U_p}} \tag{S12}$$

in which we calculate $U_p$ as follows(*18*),

$$U_p = \frac{e^2 \beta^2 E^2}{4 m_e \omega^2} \tag{S13}$$

To demonstrate most of our experiment is performed under the field-driven ($\gamma>1$) condition. We calculated $\gamma$ under a fixed focus spot size and pulse duration, as shown in Fig. S10.

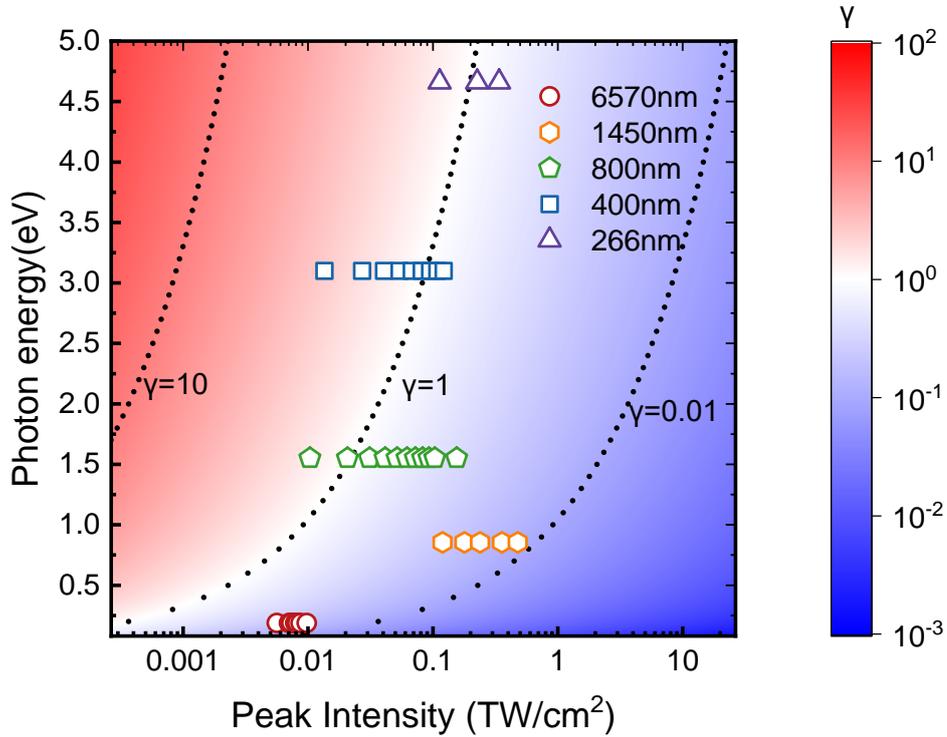

Fig. S10 Keldysh parameter versus peak fluence for various photon energies. For 400 nm (blue square dot), the maximum (left) of $\gamma$ is 2.54.

## S10. Modeling of Bias-dependent nonlinearity.

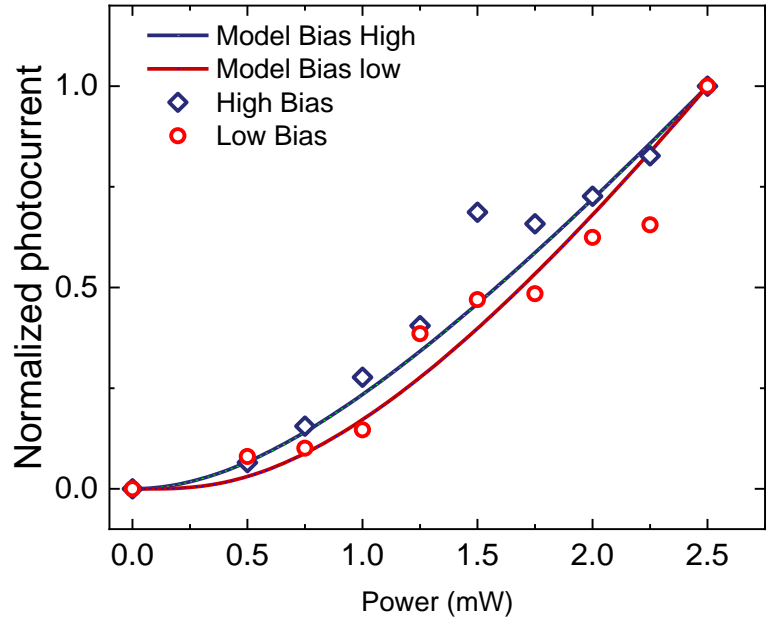

Fig. S11 Power-photocurrent relationship under different bias. Comparing to peak optical field, DC field is relatively weak (~0.01V/A), however, a slight slope change could be observed. We averaged 2.5V-3V and 0-0.5V photocurrent-power slope, and the experimental data showed a predicted trend from F-E model.

## S11.  Modelling photocurrent generation based on F-N tunneling and classic trajectories of electrons.

In PAN, electron pulse trains are generated by the optical field, with electrons emitted at various moments within the optical cycle exhibiting different kinetic energies. As the electron beam travels from one electrode to another, it experiences time delays and stretching. While precise measurements are required to determine the electron pulse shape, methods such as electron-photon cross-correlation techniques are applicable. However, the severity of delays and deformations can be estimated through calculations of electron trajectories. These methods include the Time-Dependent Schrödinger Equation (TDSE) and the classical Simple-Man Model (SMM), both of which have been successful in predicting the shape of the kinetic spectrum (34). Computational estimates indicate that, due to the relatively small nanogap distance, the pulse delays are on the order of femtoseconds, and the electron pulses do not exhibit significant broadening. The rate of electron emission is calculated using Fowler-Nordheim (F-N) emission, and their trajectory is determined using classical kinetics:

$$a(t) = -\frac{e\gamma E}{m_e}$$

Taking a similar form from tip enhancement on one-electrode field emission, we put exponential decay on both sides of the electrode with the following form:

$$\gamma = \sum_{i=1}^{2}(\alpha - 1) * \left(\frac{r_i}{r_0 + x}\right)^3 + \exp\left(-2ln2 * \frac{(x - x_{center})^2}{d_{focus}^2}\right)$$

Fig. S11 illustrates the density of electrons emitted at various phases of the optical field. Electrons emitted during the tail of the pulse or at times when the optical field strength within a cycle is weak exhibit a "quiver motion." This occurs because the field strength is insufficient to propel the electrons across to the opposite electrode within half a cycle. Nonetheless, due to the enhancement profile, the electrons do not decelerate as much as they accelerate, allowing most of them to continue moving forward when the next positive half-cycle begins. A small portion of the electrons will return to the same electrode, where they are considered to be absorbed. We did not account for elastic collisions in this process. This dynamic can lead to pulse distortion. Due to relatively short gap distances, it is possible to maintain the duration of the electron pulse at a level comparable to that of the optical pulse, as depicted in Fig. S12.

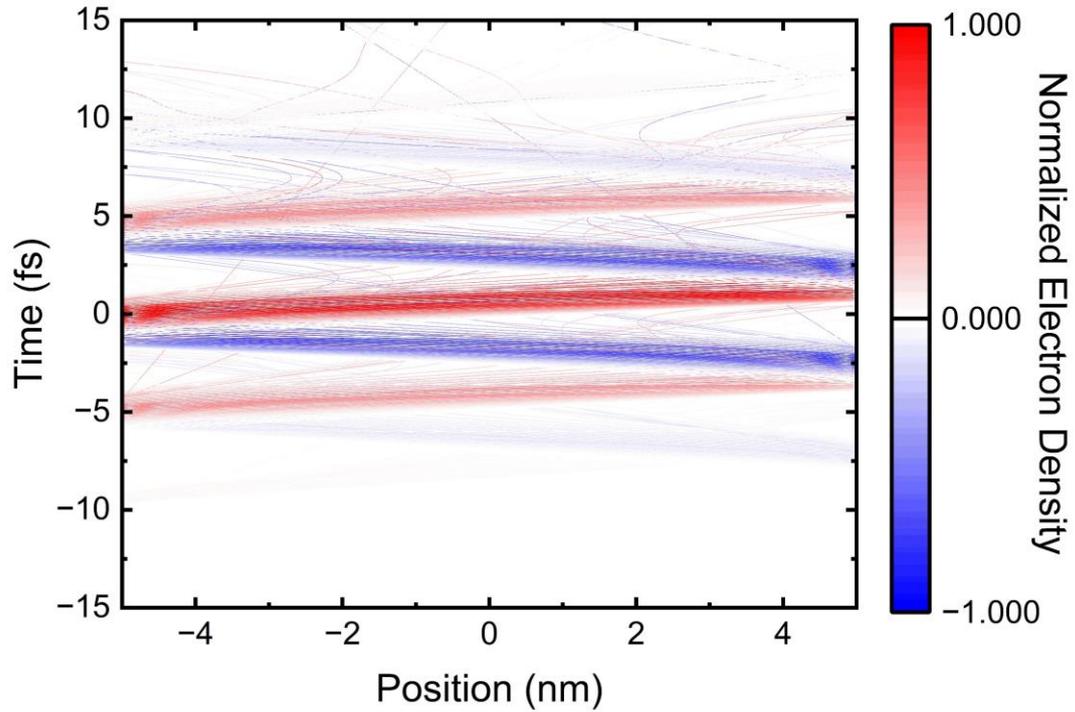

Fig. S12 Electron trajectory analysis under 1450 nm, 10 fs illumination reveals the maximum electron density emitted from Aluminum (red) and gold (blue). Most electrons arrive at the opposite electrode after a short delay. Depending on the phase in the optical-cycle, several electrons exhibit quiver motion, and some return to the emitting electrode.

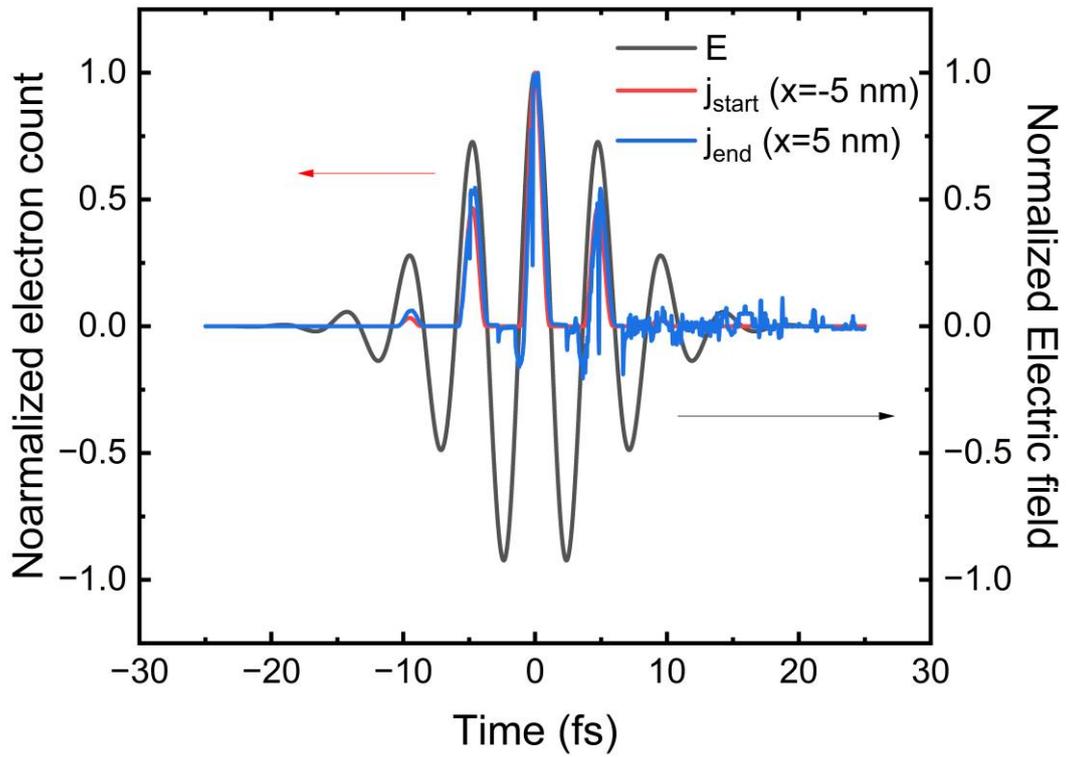

Fig. S13 Deformation of the electron pulse occurs under 1450 nm, 10 fs illumination. We analyzed the initial temporal profile of electrons emitted from Aluminum (red) and the final temporal profile upon arrival at the gold electrode (blue). This delay increases as the ionization wavelength shifts to shorter wavelengths, corresponding to shorter optical cycles.

## S12. Characterization of 800nm chirped pulses using PANs

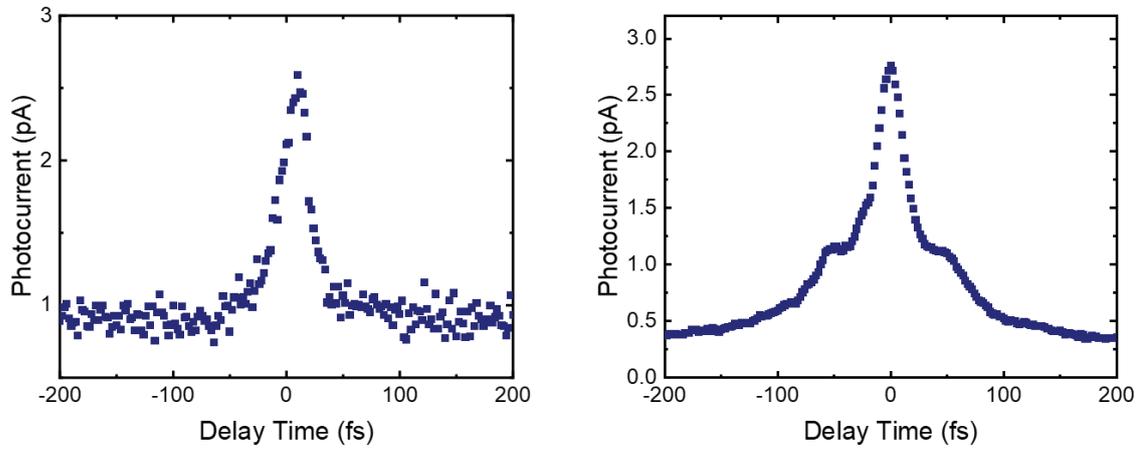

Fig. S14 Characterization of 800nm chirped pulses using nanogap. a) without chirping and b) with chirping.

Intensity autocorrelation is performed on a compressed 800 nm pulse by changing its compression. Nanogap autocorrelation can clearly follow the distortion of the pulse shape introduced by the intentional chirping.

## S13. Spectrum retrieved from MIR autocorrelation.

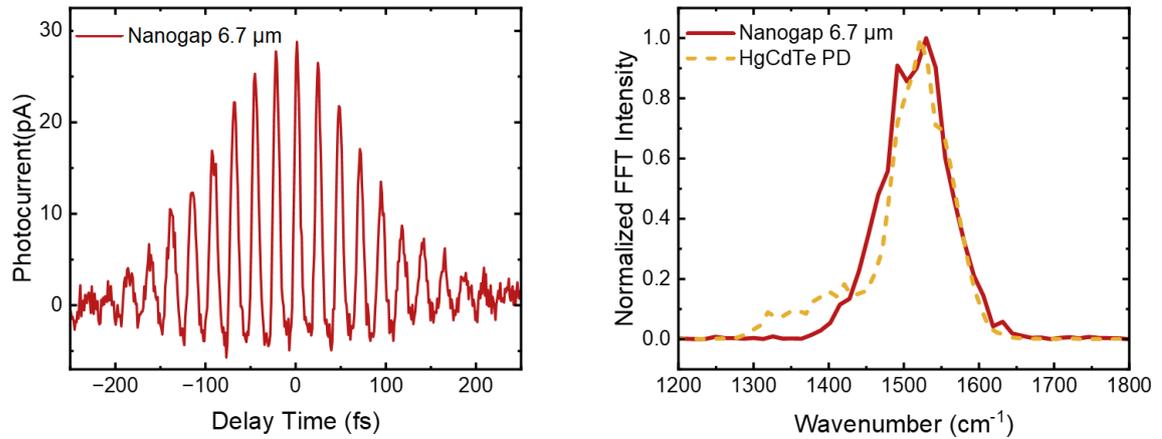

Fig. S15 MIR photocurrent autocorrelation from PAN. Left, interferometric photocurrent for collinear autocorrelation with 6.7 μm pulses. Right, Retrieved spectrum comparison between Pan autocorrelation and linear response of HgCdTe detector.

The collinear autocorrelation for the case of MIR 6.7μm gives interferometric fringes, allowing Fourier Transform to extract the MIR spectrum. The spectrum extracted from the PAN is compared with the spectrum extracted from the MIR linear response of HgCdTe detector.

## S14. 400nm autocorrelation.

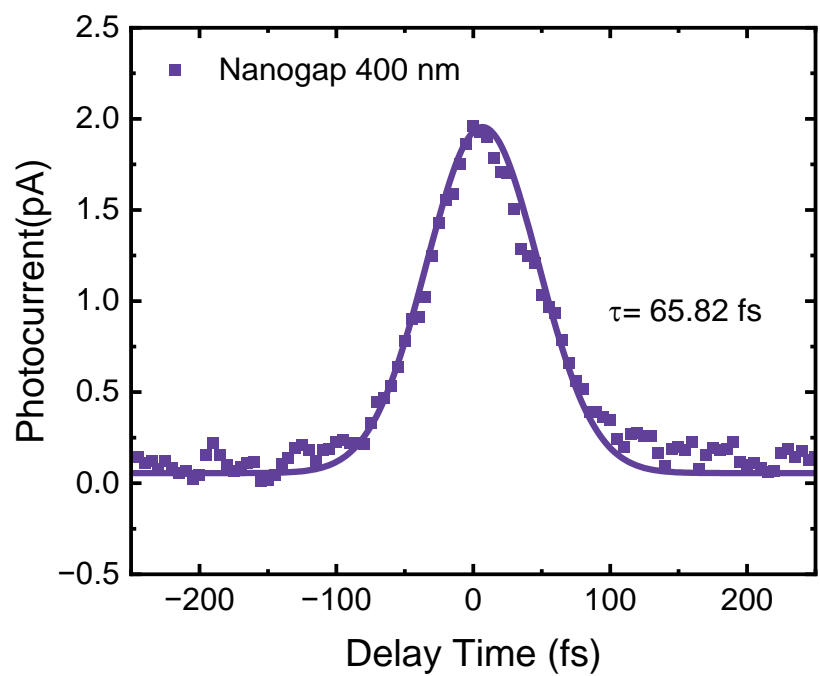

Fig. S16 Autocorrelation photocurrent on PAN for 400nm pulse.

## S15. Simulation of autocorrelation in PAN

In an $n^{th}$ order autocorrelation, Intensity of an n-photon response crystal/photodetector gives a photocurrent according to:

$$I(\tau) = \int_{-\infty}^{\infty} |[E + E(t-\tau)]^2|^n dt \qquad (S14)$$

In PAN, we could have a correlation according to instant nonlinear response from a similar term,

$$J(\tau) = \int_{-\infty}^{\infty} j(E + E(t-\tau)) \, dt \qquad (S15)$$

With a sufficiently strong field strength where the slope is slowly varying, it has a feature similar to fixed order autocorrelation, which is compared in Fig. S17.

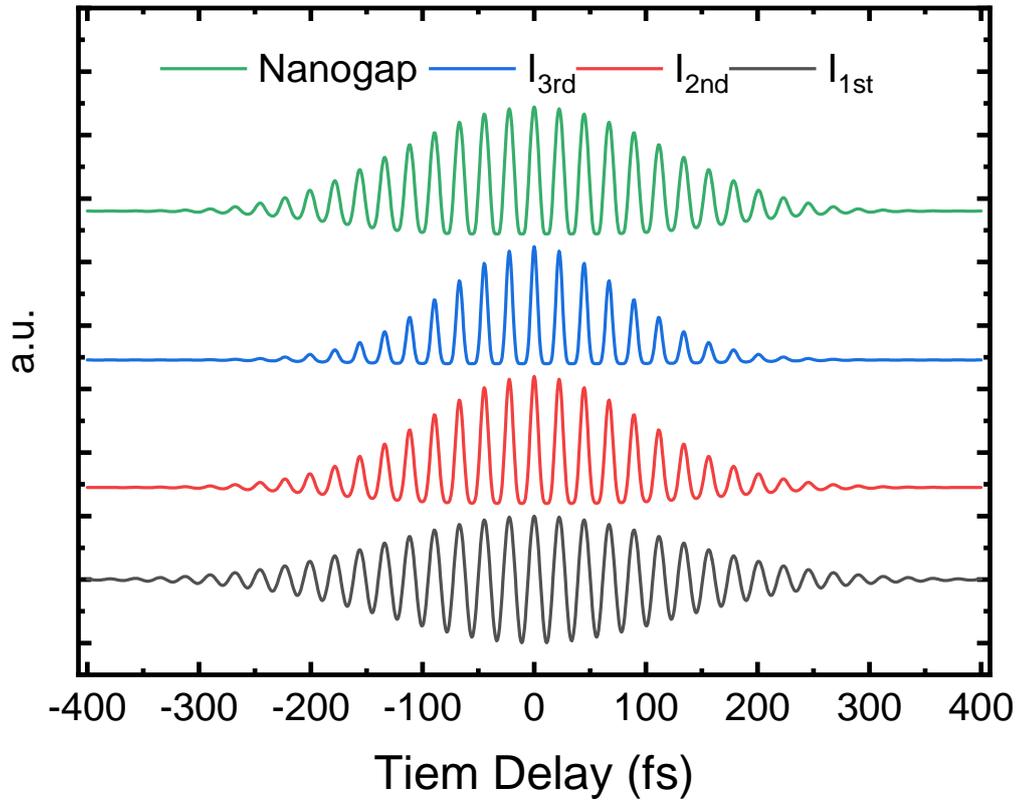

Fig. S17 Comparison of calculated nanogap autocorrelation and various order of nonlinear response. Interferometric autocorrelation is calculated with a center wavelength of 6.7μm.

# S16. Fabrication of QD and ZnO devices based on PANs

**Chemicals:**

Cesium Bromide (CsBr, 99.999%), Lead bromide (PbBr$_2$, 99.999%), dibenzo-21-crown-[7] (DB21C7, 97%), N,N-dimethylformamide (DMF, anhydrous, ≥99.9%), Toluene (anhydrous, ≥99.5%), oleic acid (OA, ≥99%), octylamine (OctAm, ≥98%), n-butanol (anhydrous, ≥99.8%), octadecylphosphonic acid (ODPA, 97%), zinc oxide (ZnO, 99.99% trace-metal basis), ammonium hydroxide (50% v/v aqueous). All chemicals were purchased from Sigma-Aldrich and used as received.

**Synthesis of CsPbBr3 QD Nanocubes:**

Precursor solutions were prepared by adding 0.1 mmol of CsBr, 0.1 mmol of DB21C7, 0.1 mmol of PbBr2, and 1 ml of DMF in a 7 ml vial under stirring until all chemicals were dissolved. Subsequently, 150 μl of the precursor was added dropwise into a vigorous stirring solution containing 5 ml of toluene, 0.3 ml of OA, 25 μl of OctAm, and 2 ml of n-butanol for 1 min. Following this, the crude reaction solution was purified using two centrifuge steps.

In the first step, the crude reaction solution was centrifuged at 8500 rpm for 10 mins. The precipitant was collected and redispersed in 1 ml toluene. The dispersion was then centrifuged at 3000 rpm for 1 min. The supernatant was collected and filtered through a 0.22 um filter before deposition onto the nanogap electrode chip.

To confirm that the synthesized quantum dots meet the expected size and are sufficiently small to enter nanogaps, we characterized them through both transmission electron microscopy (TEM) and UV-VIS absorption/PL spectroscopy. We analyzed the size statistics from TEM (Fig. S18) and the position of the PL peaks (Fig. S19). Both results indicated that the size of the majority of quantum dots meets the requirements for entry into nanogaps.

**Preparation of ZnO solution:**

Precursor solution of ZnO was prepared by dissolving zinc oxide powder in ammonium hydroxide at a concentration of 10 mg ml$^{-1}$.

## Deposition of ZnO and CsPbBr3 QDs into the nanogap:

### ZnO:

The ZnO solution was spin-coated onto the nanogap electrodes at 4000 rpm for 40 s, followed by 35min thermal annealing at 180°C in air. A second layer of the same solution was spin-coated, following exactly the same procedure to yield a film thickness of around 40nm.

### CsPbBr3 QD:

The 1 mL of prepared QDs solution was diluted with 10 mL of toluene. Then, 100 μL of the diluted QDs solution and 1 mL of toluene were added into a 7 mL vial. A chip was soaked in the mixture for 7 hours, after which it was picked up and dried for 30 minutes. The dried chip was used for further measurements. A clear J-V characteristic difference can be seen among the junction with the SAM layer, the empty junction, and the junction with QDs (Fig. S20, Fig. S21).

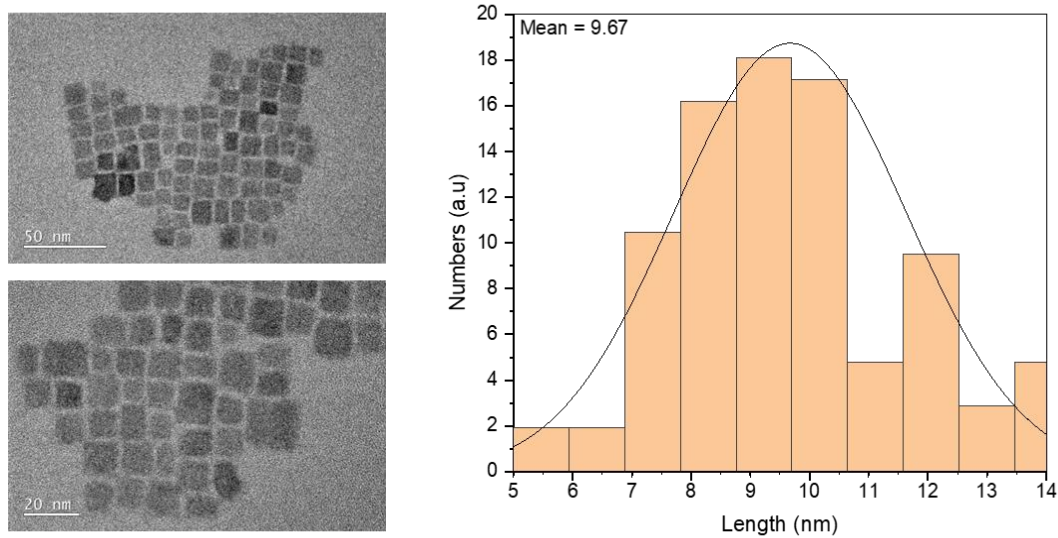

Fig. S18 TEM images of synthesized CsPbBr$_3$ nanocrystals and the size statistics.

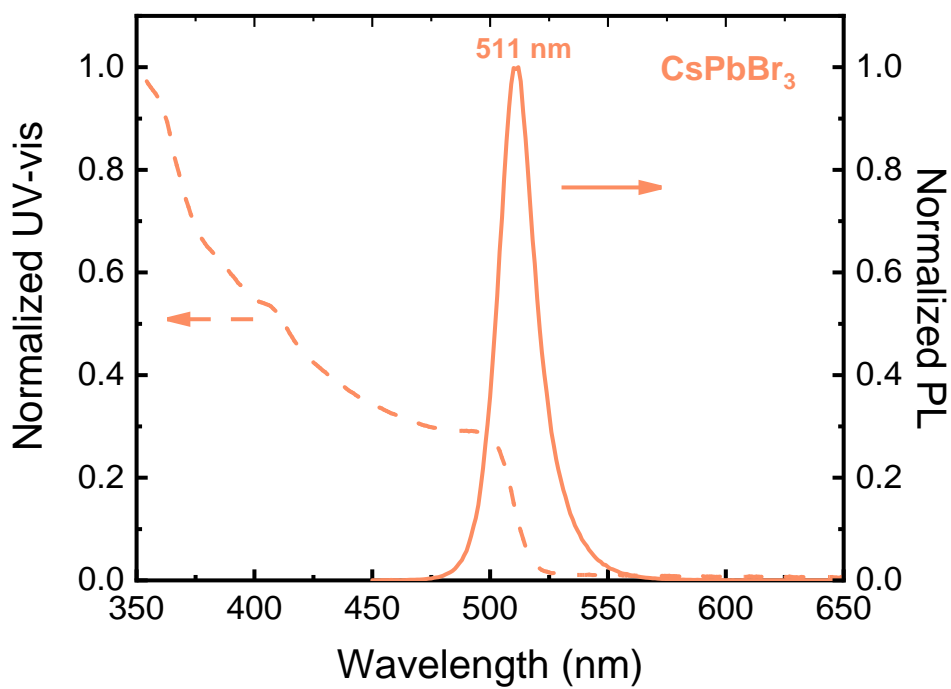

Fig. S19 Absorption and PL spectra of CsPbBr$_3$ nanocrystals with <10nm average size. The excitonic feature is seen in the absorption part, and the PL peak position indicates the average size is between 9 nm~10 nm. FWHM of PL peak is around 90 meV.

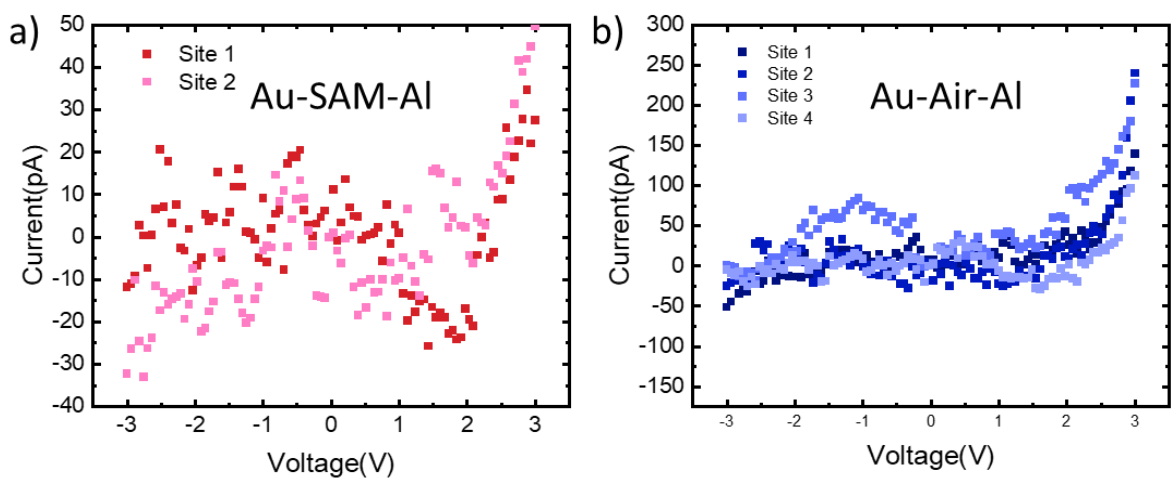

Fig. S20 Implementation of perovskite QDs into the nanogap. I-V characteristic curves for various states of the nanogap: a) air, including the SAM-layer on the Al electrode, b) air, after removing the SAM.

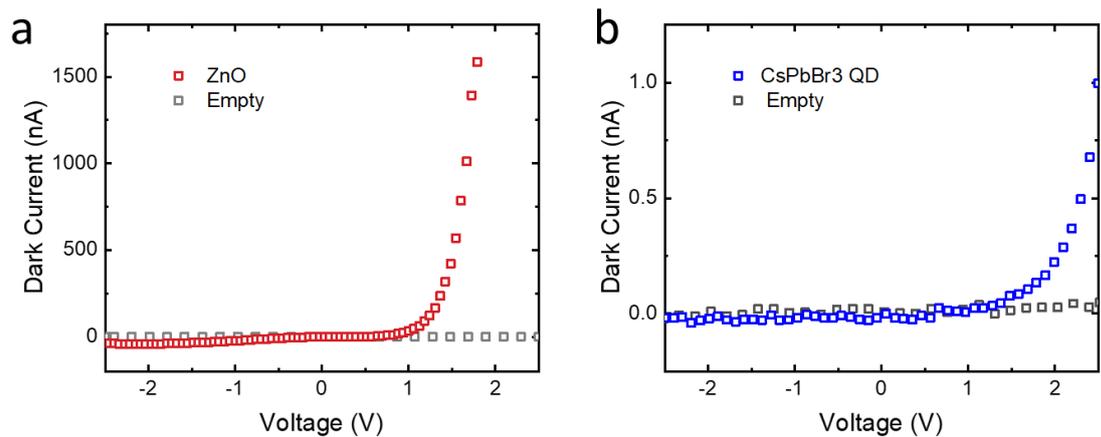

Fig. S21 I-V characteristics of different materials within the nanogap. Black dots indicate a reference from the air gap. a) ZnO. b) CsPbBr3 QD. ZnO shows a much higher current due to its very low conduction band, which is closer to the work function of the Al electrode.

## S17. Evidence of negligible optically induced current in the glass substrate

It has been demonstrated that carriers can be injected from dielectric material to the electrode to generate photocurrent(*2*). Therefore, apart from carrier generation via F-N emission, carrier injection from the glass substrate in PANs can be a second origin of photocurrent under a strong field. However, our results show that the F-N emission mechanism is the dominant pathway of photocurrent generation in PANs for the following reasons.

Most of our measurements are performed under a peak field strength around 0.1 V/Å, which is 20 times smaller than the working peak field strength (400 times smaller on fluence) to inject electrons from the dielectric (2 V/Å). Therefore, the peak field strength in our measurements is insufficient to inject electrons from the glass substrate. Also, the dielectric between two electrodes in PAN is only 10nm, which further reduces the possible injected electrons.

In the traditional symmetrical device structure, both electrodes have the same work function in the absence of applied voltage. Although the threshold of electron ionization from the metal (gold, ~5.1 eV) is smaller than the dielectric (e.g. for silica is ~11 eV), the electrons in both metal electrodes are ionized with a similar possibility, resulting in similar photocurrent canceling throughout full optical cycles. Therefore, the major contribution to the detectable photocurrent is obtained from electron injection from the dielectric material rather than the electrons ionized from the metal electrode. However, in PANs, we use asymmetrical device structures (Al-Au) with different work functions ($\emptyset_{Al} \approx 4.3\ eV$ and $\emptyset_{Au} \approx 5.2\ eV$) to enable different possibility of electron ionization from the two different metal electrodes without the assistance of applied voltage. This asymmetric electrode design allows us to observe detectable photocurrent with a small peak field strength (i.e., fluence).